\documentclass[conference]{IEEEtran}
\usepackage{amsmath,amsfonts,amssymb}

\usepackage[noend]{algpseudocode}
\usepackage{algorithm}

\usepackage{array}
\usepackage[caption=false,font=normalsize,labelfont=sf,textfont=sf]{subfig}
\usepackage{textcomp}
\usepackage{stfloats}
\usepackage{url}
\usepackage{verbatim}
\usepackage{graphicx}
\usepackage{cite}
\usepackage{mathptmx} 
\usepackage{xcolor}
\usepackage{import}
\usepackage{enumitem}
\usepackage{subcaption}
\usepackage{booktabs}
\usepackage[acronym,nomain]{glossaries}
\usepackage{multirow}
\usepackage[bottom]{footmisc}
\usepackage{float}
\usepackage{subfloat}
\usepackage{tikz}
\usepackage{calc}
\usepackage{xfrac}
\usepackage{color,soul}
\usepackage{enumitem}

\usepackage{wrapfig}
\usepackage{multicol}
\usepackage{lmodern}
\usepackage{caption}
\makeglossaries 

\newglossaryentry{latex}
{
        name=latex,
        description={Is a mark up language specially suited for
scientific documents}
}

\newglossaryentry{bibliography}
{
        name=bibliography,
        plural=bibliographies,
        description={A list of the books referred to in a scholarly work,
typically printed as an appendix}
}

\newglossaryentry{maths}
{
    name=mathematics,
    description={Mathematics is what mathematicians do}
}

\newacronym{fg}{FG}{forwarding graph}
\newacronym{nbi}{NBI}{northbound interface}
\newacronym{qoe}{QoE}{quality of experience}
\newacronym{ibn}{IBN}{intent-based networking}
\newacronym{rq}{RQ}{research question}
\newacronym{ro}{RO}{research objective}
\newacronym{sfc}{SFC}{service function chain}
\newacronym{mano}{MANO}{network functions virtualization management and orchestration}
\newacronym{nfv}{NFV}{network function virtualization}
\newacronym{nfvo}{NFVO}{\gls{nfv} orchestrator}
\newacronym{sdnc}{SDNC}{\gls{sdn} controller}
\newacronym{ems}{EMS}{element management system}
\newacronym{eca}{ECA}{event-conditon-action}
\newacronym{hmi}{HMI}{human-machine interaction}
\newacronym{anima}{ANIMA}{autonomic networking integrated model and approach}
\newacronym{wg}{WG}{working group}

\newacronym{vnf}{VNF}{virtualized network function}
\newacronym{pnf}{PNF}{physical network function}
\newacronym{nfvi}{NFVI}{network function virtualization infrastructure}
\newacronym{tmn}{TMN}{telecommunications management network}
\newacronym{ran}{RAN}{radio access network}
\newacronym{cn}{CN}{core network}
\newacronym{pbnm}{PBNM}{policy-based network management}
\newacronym{sdn}{SDN}{software-defined network}
\newacronym{qos}{QoS}{quality of service}
\newacronym{nmrg}{NMRG}{network management research group}
\newacronym{etsi}{ETSI}{european telecommunications standards institute}
\newacronym{ssot}{SSoT}{single source of truth}
\newacronym{dsl}{DSL}{domain-specific language}

\newacronym{eni}{ENI}{experiential network intelligence}
\newacronym{zsm}{ZSM}{zero-touch network and service management}

\newacronym{sla}{SLA}{service-level agreement}
\newacronym{slo}{SLO}{service-level objective}
\newacronym{sls}{SLS}{service-level specification}
\newacronym{nl}{NL}{natural language}
\newacronym{cnl}{CNL}{constrained natural language}

\newacronym{nlp}{NLP}{natural language processing}
\newacronym{sdo}{SDO}{standards development organization}
\newacronym{tmf}{TMForum}{Telemanagement Forum}
\newacronym{owl}{OWL}{web ontology language}
\newacronym{rdf}{RDF}{resource description framework}
\newacronym{rdfs}{RDFS}{resource description framework schema}
\newacronym{iri}{IRI}{internationalized resource identifier}

\newacronym{api}{API}{application programming interface}
\newacronym{kg}{KG}{knowledge graph}
\newacronym{kge}{KGE}{knowledge graph embedding}

\newacronym{kb}{KB}{knowledge base}

\newacronym{kpi}{KPI}{key performance indicator}
\newacronym{cli}{CLI}{command line interface}
\newacronym{snmp}{SNMP}{simple network management protocol }
\newacronym{itu}{ITU}{international telecommunication union }
\newacronym{5g}{5G}{fifth generation}
\newacronym{nsa}{NSA}{non-standalone}
\newacronym{gnb}{gNB}{gNodeB}
\newacronym{mse}{MSE}{mean squared error}

\newacronym{e2e}{E2E}{end-to-end}
\newacronym{3gpp}{3GPP}{3rd generation partnership project}

\newacronym{csp}{CSP}{communication service provider}
\newacronym{csc}{CSC}{communication service customer}
\newacronym{nop}{NOP}{network operator}
\newacronym{nip}{NIP}{network infrastructure provider}
\newacronym{rfc}{RFC}{request for comments}

\newacronym{sba}{SBA}{service-based architecture}
\newacronym{csmf}{CSMF}{communication service management function}
\newacronym{cfcs}{CFCS}{customer facing communication service}
\newacronym{rfcs}{RFCS}{resource facing communication service}
\newacronym{dsd}{DSD}{dynamic service descriptor}
\newacronym{lcm}{LCM}{lifecycle management}
\newacronym{csi}{CSI}{communication service instance}
\newacronym{nsi}{NSI}{network slice instance}
\newacronym{nssi}{NSSI}{network slice subnet instance} 
\newacronym{nsmf}{NSMF}{network slice management function}
\newacronym{oam}{OAM}{operation and maintenance}
\newacronym{mc}{MC}{mission-critical}
\newacronym{nmc}{NMC}{non-mission-critical}
\newacronym{ns3}{ns3}{network simulator 3}
\newacronym{cl}{CL}{closed loop}
\newacronym{ai}{AI}{artificial intelligence}
\newacronym{ml}{ML}{machine learning}
\newacronym{nsd}{NSD}{network service descriptor}
\newacronym{rl}{RL}{reinforcement learning}
\newacronym{yaml}{YAML}{yet another markup language}

\newacronym{ietf}{IETF}{Internet Engineering Task Force}

\newacronym{embb}{eMBB}{enhanced mobile broadband}
\newacronym{urllc}{uRLLC}{ultra-reliable low-latency communication}
\newacronym{mmtc}{mMTC}{massive machine type communication}

\newacronym{xr}{XR}{extended reality}
\newacronym{vr}{VR}{virtual reality}
\newacronym{ar}{AR}{augmented reality}
\newacronym{fcaps}{FCAPS}{fault, configuration, accounting, performance and security}
\newacronym{dikw}{DIKW}{Data-Information-Knowledge-Wisdom}
\newacronym{osm}{OSM}{open-source \gls{mano}}
\newacronym{sdwan}{SD-WAN}{software-define wide area network}
\newacronym{rest}{REST}{representational state transfer}
\newacronym{nwdaf}{NWDAF}{network data analytics function}
\newacronym{vnfd}{VNFD}{virtualized network function descriptor}
\newacronym{mape}{MAPE}{(Monitor-Analyze-Plan-Execute over a shared Knowledge)}
\newacronym{ooda}{OODA}{observe orient decide act}
\newacronym{mracl}{MRACL}{model reference adaptive control loop}
\newacronym{focale}{FOCALE}{foundation observe compare act learn reason}
\newacronym{den}{DEN}{directory enabled network}
\newacronym{denon}{DENON}{directory enabled network ontology}
\newacronym{gana}{GANA}{generic autonomic networking architecture}
\newacronym{onix}{ONIX}{overlay network for information exchange}

\newacronym{b2b}{B2B}{business-to-business}
\newacronym{b2h}{B2H}{business-to-household}
\newacronym{b2c}{B2C}{business-to-consumer}
\newacronym{b2b2x}{B2B2X}{business-to-business-to-everything}

\newacronym{netconf}{NETCONF}{network configuration protocol}

\newacronym{ge}{GE}{Gaussian embedding}
\newacronym{pbm}{PBM}{policy based management}
\newacronym{onos}{ONOS}{open network operating system}
\newacronym{icm}{ICM}{intent common model}
\newacronym{mgda}{MGDA}{Multiple Gradient Descent Algorithm}
\newacronym{spsa}{SPSA}{Stochastic Perturbation Stochastic Approximation}
\newacronym{sgd}{SGD}{stochastic gradient descent}

\newacronym{pdb}{PDB}{packet delay budget}
\newacronym{ptt}{PTT}{push-to-talk}
\newacronym{mr}{MR}{mean rank}
\newacronym{aqm}{AQM}{active queue management}
\newacronym{fq-codel}{FQ-CoDel}{FlowQueue-Controlled Delay}
\newacronym{kgl}{KGL}{knowledge graph learning}
\newacronym{kg2e}{KG2E}{knowledge graph with Gaussian embedding}

\newacronym{irtf}{IRTF}{internet research task force}

\newacronym{mapek}{MAPE-K}{(monitor-analyze-plan-execute over a shared knowledge}
\newacronym{ner}{NER}{named entity recognition}

\newacronym{tpr}{TPR}{true positive rate}
\newacronym{tnr}{TNR}{true negative rate}
\newacronym{fpr}{FPR}{false positive rate}
\newacronym{fnr}{FNR}{false negative rate}

\newacronym{tp}{TP}{true positive}
\newacronym{tn}{TN}{true negative}
\newacronym{fp}{FP}{false positive}
\newacronym{fn}{FN}{false negative}

\newacronym{ikg}{IKG}{intent knowledge graph}

\usepackage [english]{babel}
\usepackage [autostyle, english = american]{csquotes}
\MakeOuterQuote{"}

\newcommand{\tilda}{~}
\begin{document}
%
\title{Knowledge Graph Embedding in Intent-Based Networking}

\author{\IEEEauthorblockN{Kashif Mehmood, Katina Kralevska, and David Palma} \\ \vspace{-0.3cm}
\IEEEauthorblockA{\textit{Department of Information Security and Communication Technology (IIK)} \\
\textit{NTNU\textemdash Norwegian University of Science and Technology, Trondheim, Norway}\\
Email: \{kashif.mehmood, katinak, david.palma\}@ntnu.no}
}

\maketitle


%
\IEEEpeerreviewmaketitle


\begin{abstract}
This paper presents a novel approach to network management by integrating \gls{ibn} with \glspl{kg}; thus, creating a more intuitive and efficient pipeline for service orchestration. By mapping high-level business intents onto network configurations using \glspl{kg}, the system dynamically adapts to network changes and service demands, ensuring optimal performance and resource allocation. This integration facilitates a deeper understanding of network states and dependencies, enabling predictive adjustments and real-time troubleshooting. We utilize \gls{kge} to acquire context information from the network and service providers. The trained \gls{kge} model maps intents to services via service prediction and intent validation processes in the proposed intent processing pipeline. We evaluate the trained model for its efficiency in the service mapping and intent validation tasks using simulated environments and extensive experiments. The service prediction and intent verification accuracy $\geq 80\%$ is achieved for the trained \gls{kge} model on a custom service orchestration \gls{ikg} based on TMForum's intent common model.
\end{abstract}

\section{Introduction}
Network and service providers face the daunting task of supporting applications and user demands at massive scales, all while ensuring compliance with essential business and operational requirements such as security, availability, and latency~\cite{intro-fuqaha-iot}. Traditional network management systems are often time-consuming and error-prone due to manual configurations and maintenance procedures~\cite{intro-kreutz-sdn}. Moreover, managing networks in multi-domain environments, characterized by heterogeneous resources with varying abstractions, introduces complexities and potential incompatibilities. To address these challenges, changes in network architecture are underway, leveraging automation techniques such as software-defined networking and network function virtualization~\cite{MEHMOOD2023109477, ibn-survey-leivadeas}.\par

\gls{ibn} has been met with significant interest as a pioneering approach to network automation and service orchestration aiming to alleviate these challenges. The \gls{irtf} \gls{nmrg} has made strides toward establishing a standardized understanding of "intent" within the IBN framework~\cite{intentdefsrfc9315}. They describe an intent as a collection of operational goals and outcomes that a network endeavors to achieve, articulated declaratively. This definition, however, stops short of specifying the means or methods for actualizing these goals and outcomes. \gls{ibn}'s core philosophy enables stakeholders, such as network administrators, application developers, network operators, and subscribers, to define their desired network behavior using a high-level language, bypassing the complexities of low-level network configurations. \Gls{irtf} specifies the following phases as the core part of an intent's life cycle: i) \textit{user profiling and expression}, ii) \textit{modeling and translation}, iii) \textit{planning and conflict resolution}, iv) \textit{activation and deployment}, and v) \textit{assurance}~\cite{intentdefsrfc9315}. This work focuses on intent modeling and translation. It utilizes the available context using \glspl{kg} in the network domain to understand the intent expression and pave the path for its activation in the underlying network infrastructure.\par

The integration of \gls{rdf} and \gls{owl} into the modeling of \gls{ibn} represents a pivotal advancement in the pursuit of intelligent, adaptable, and semantically rich network management systems. By leveraging these technologies, the development of detailed knowledge graphs that facilitate a nuanced understanding of network dynamics, policies, and, eventually, intents is made possible. By modeling networks as knowledge graphs, operators can understand the network's current state and how it relates to the desired operational state dictated by an intent. This semantic richness enables more precise and dynamic mapping of intents to network configurations, significantly improving the network's adaptability and responsiveness to changing network conditions.\par

\textit{Randles et al.}~\cite{kg-ont-cl-ibn} propose an ontology for mapping the intent flow in \gls{cl} automation framework. Inspired by the MAPE-K (Monitor-Analyze-Plan-Execute over a shared Knowledge) framework, an intent ontology describes different components such as metrics, action, goals. \textit{Dzeparoska et al.}~\cite{ibn-kg-kb} utilize a knowledge base of known intents and their actions to query and train the intent translation model for network policy generation. \textit{Wang et al.}\tilda\cite{kg-vert-ibn} explore different sources of information in a vertical industry that can be incorporated in a \gls{kg} and propose a \gls{kg} construction scheme using these sources of information. \textit{Daroui et al.}\tilda\cite{ibn-kg-state} focus on developing a distributed knowledge base concept concerning different deployable intents in the network and propose a solution for state management of these distributed knowledge bases. However, there is still a lack of literature regarding integrating \glspl{kg} in the intent life cycle for aiding intent translation and deployment phases.

The contributions of this paper are as follows:
\begin{itemize}
    \item We design a knowledge graph storage solution by embedding the available \gls{rdf} triples as projections in Gaussian space. This operation provides a trained \gls{kge} model for the completion and verification of intents during the intent processing and translation.
    \item We propose an intent processing pipeline with active reporting and planning modules to adapt to the changing network state during the life cycle of a deployed intent. The pipeline uses the proposed \gls{ikg} and a trained \gls{kge} model to map intents to respective services for deployment. 
    \item For validation of the proposed pipeline, we design and analyze extensive experimental campaigns covering several boundary conditions and challenges encountered in the simulated environment. 
\end{itemize}

The rest of the paper is organized as follows: Section II describes the proposed pipeline's components and their integration with knowledge. Section III presents the evaluation results. Section IV concludes the paper.

\section{Proposed Intent Processing Pipeline with Knowledge Graph Embedding}
This section first explains the key components of the proposed pipeline --- intent processing and \gls{kge} learning. Next, it presents the proposed pipeline for integrating context as a knowledge graph in the intent life cycle. The relationships between intents, services, and associated \glspl{kpi} follow the service orchestration concept in 5G and B5G networks.


\subsection{Integration of Knowledge Graph in Intent Processing}
\begin{figure}[!htbp]
\begin{center}
    \includegraphics[width=0.49\textwidth]{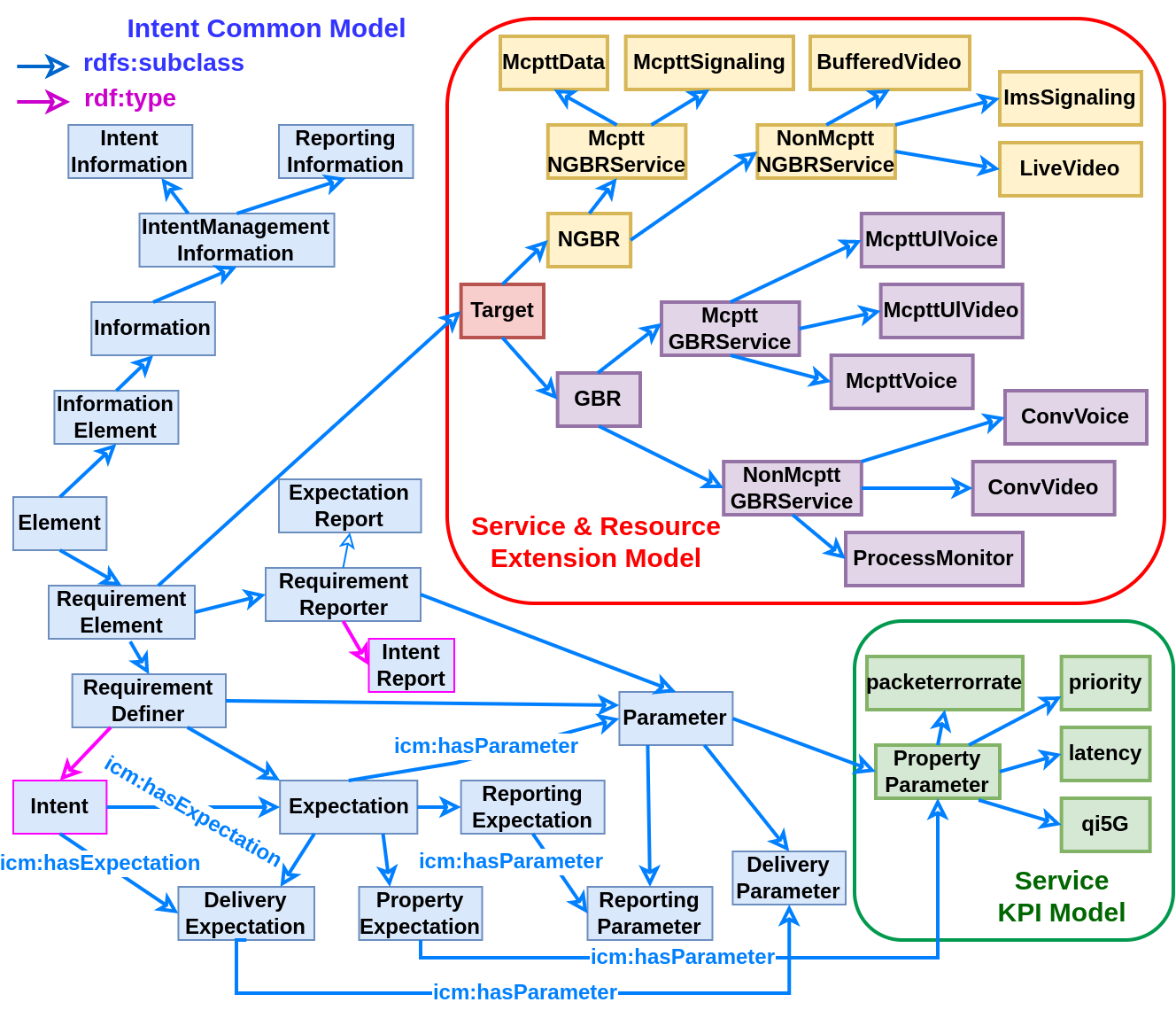}
\end{center}
\caption{An Intent Knowledge Graph (IKG) for \gls{ibn}\tilda\cite{ibn-kg-self}.}
\label{image: ibn-kg-combined}
\end{figure}

\Glspl{kg} consist of triples describing facts arranged in a systematic language using a well-known structure such as \gls{owl}~\cite{owl2-w3c} and \gls{rdf}~\cite{rdf-w3c}. Intents also have a specific structure with information fields describing a particular characteristic of the application use case. For example, \textit{"IntentA has an expectation Exp1"} is represented as the triple $(IntentA, hasExpectation, Exp1)$. In this way, any data object can be represented using a \gls{kg} in the form of triples. Hence, we can model an intent as a \gls{kg} with different entities and relationships between them as depicted in the \gls{icm} by \gls{tmf}~\cite{tmforum-intent-common-model-tr290}. In this paper, we utilize the \gls{ikg}, given in \figurename\tilda\ref{image: ibn-kg-combined}, that is based on an extension of the \gls{tmf}'s \gls{icm}\tilda\cite{ibn-kg-self}.

The service, resource, and KPI extension models in the IKG are essential components for defining service requirements in a service template. These components in the IKG represent information relevant to the available service offerings, resource types, and associated KPIs for the defined services. For example, a triple \textit{(GBR, rdfs:subclass, NonMcpttGBRService)} represents a subclass of guaranteed bit rate (GBR) resource type, i.e., non-mission critical GBR service. Furthermore, the triple \textit{(PropertyParameter, rdfs:subclass, latency)} represents a KPI parameter known as `latency' for the defined services.

\subsubsection{A Simple Knowledge Graph Model}
\begin{figure}[!htbp]
\begin{center}
    \includegraphics[width=0.49\textwidth]{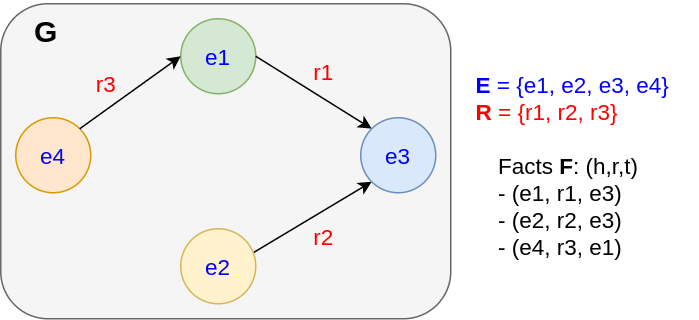}
\end{center}
\caption{An example of a \gls{kg}.}
\label{image: kg-simple}
\end{figure}

A \gls{kg} $\mathcal{G}$ is described as the following formulation, 

\begin{equation} \label{eq:1}
    \mathcal{G} = {(E, R, F)}
\end{equation}

where $E = \{e_1, e_2, e_3, ..., e_{|E|}\}$  is the set of entities (subject or object), $R = \{r_1, r_2, r_3, ..., r_{|R|}\}$ is the set of relations, and $F  \subseteq E \times R \times E$ is the set of fact triples and each triple is denoted as $(h,r,t)$. Here, $h$ and $t$ are head and tail entities, and $r$ denotes their relation. An entity is an object that can be classified as a class, type, or literal. Class objects utilize the \gls{rdfs} to differentiate types of entities as distinct classes in a KG. Type object is used to identify class relationships as \textit{rdf:type} objects; for example, class1 and class2 have similar properties, and they are of the same type, that is, class A. A literal object is based on \textit{rdf:literal} objects containing specific values as strings or integers. An example of a \gls{kg} is depicted in \figurename\tilda\ref{image: kg-simple}. \par

\subsubsection{Modeling Intents from IKG}
In the \gls{ikg}, one of the intent triples is \textit{(icm:PropertyExpectation, icm:hasParameter, icm:PropertyParameter)} representing the relationship between an expectation and its associated parameter (i.e., "\textit{PropertyExpectation has a Parameter that is PropertyParameter1}"). Here, the intent triple is complete and utilized to process an intent. 

In contrast, the triple \textit{(icm:Expectation, icm:hasTarget, icm:Target)} shows the relationship (i.e., "\textit{Expectation has a Target which is Target1}") between the intent expectation and the required target to be reached. For example, the intent translation process updates the information regarding the expected target and required service, and \gls{kpi} parameters are appended from the \gls{ikg} to complete the designed intent model. This is accomplished via a series of trained models using knowledge embedding\tilda\cite{ge-kg2e} to predict unknown entities in intent triples. 

\begin{figure}[!htbp]
\begin{center}
    \includegraphics[width=0.45\textwidth]{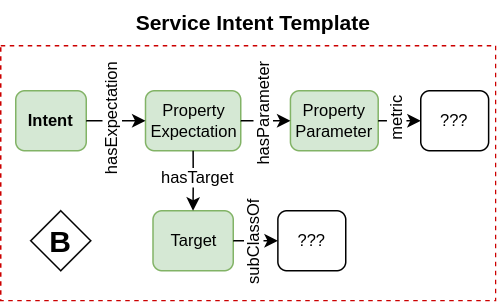}
\end{center}
\caption{An incomplete service intent template.}
\label{image: si-temp}
\end{figure}

In other words, a service intent template is defined using the \gls{ikg}, and the incomplete information, represented as `\textbf{???}' in \figurename\tilda\ref{image: si-temp}, is completed by augmenting information related to service, resource, and expected \gls{kpi} from the \gls{ikg} in \figurename\tilda\ref{image: ibn-kg-combined} using \gls{kge} learning.

\subsubsection{Knowledge Graph Embedding}
\gls{kg} analytics and reasoning are enhanced by mapping knowledge into a vector space. This involves three steps: translating entities and relations into the vector space, establishing a scoring function to evaluate their relationships, and undergoing a training process to refine the representations of entities and relations. 

Different techniques map knowledge and context from natural language to another latent space\tilda\cite{lin_knowledge_2018}. However, a \gls{ge} model effectively combines the mapping of various models such as DistMult, ComplexIE, and TransE. Moreover, the robustness, scalability, and ability to support domain-specific \glspl{kg} allow a \gls{ge}-based model to suit appropriately with the proposed \gls{ikg} and its extended version\tilda\cite{lin_knowledge_2018}. Therefore, we adopt a translational distance model based on \gls{ge}\tilda\cite{ge-kg2e}, which treats entities and relations as random vectors drawn from Gaussian distributions. This allows the model to capture entities' and relations' central tendency (mean) and uncertainties (covariance).\par

Given a triple \textit{(h,r,t)}, \gls{kg2e} models each entity and relation as a Gaussian distribution with parameters $(\mu, \mathcal{C})$, where $\mu$ is the mean vector and $\mathcal{C}$ is the covariance matrix. These vectors are distinct in their semantic meanings and vary in uncertainty compared to other entities with analogous semantics. For any given entity or relation, the average of its vector embedding marks the core of its semantic interpretation, while the covariance matrix indicates the degree of uncertainty. The distributions for entities and relations are denoted as:

\begin{equation} \label{eq:2}
    \begin{matrix}
       \textbf{h} \sim \mathcal{N}(\mu_{h}, \mathcal{C}_{h}) \\
       \textbf{r} \sim \mathcal{N}(\mu_{r}, \mathcal{C}_{r})  \\
       \textbf{t} \sim \mathcal{N}(\mu_{t}, \mathcal{C}_{t})
    \end{matrix}
\end{equation}

where $\{\mu_{h}, \mu_{r}, \mu_{t}\} \in \mathbb{R}^{d}$ and $\{\mathcal{C}_{h}, \mathcal{C}_{r}, \mathcal{C}_{t}\} \in \mathbb{R}^{d \times d} $ are mean vectors and covariance matrices, respectively.\par

\subsubsection{Training of KG Embedding Model}
The \gls{kg2e} model is trained on the \gls{ikg} to minimize the scoring function, aiming to reduce the distance between the distributions of related entities and relations. This makes the embeddings of correct triples more similar than those of incorrect triples. The scoring function $f_r (h,t)$ is defined to measure the correctness of the fact $(h, r, t)$ in the embedding space. The scoring methods in \gls{kg2e}\tilda\cite{ge-kg2e} are modeled as asymmetric Kullback-Leibler (KL)-divergence and symmetric expected likelihood, with the scoring function being categorized as translational distance-based. The scoring function is defined using the probability inner product as:
\begin{equation}\label{eq:3}
\begin{matrix}\raisetag{\baselineskip}
            f_r (h,t) = \int \mathcal{N}_{x}(\mu_{t} - \mu_{h}, \mathcal{C}_t + \mathcal{C}_h) \hspace{0.20cm} \textbf{.} \hspace{0.20cm} \mathcal{N}_{x}(\mu_{r}, \mathcal{C}_r) d\textbf{x}
    \\ \propto -\mu^{T}\mathcal{C}^{-1}\boldsymbol{\mu} - \ln(\det(\mathcal{C}))
\end{matrix}
\end{equation}

where $\boldsymbol{\mu} = \mu_{h} - \mu_{r} - \mu_{t}$ and $\mathcal{C} = \mathcal{C}_{h} + \mathcal{C}_{r} + \mathcal{C}_{t}$. Here, $\boldsymbol{\mu}$ is the difference between mean vectors and captures the relationship between the head \textit{h} and tail \textit{t} entities using relation \textit{r}. It represents the relational embedding for $(h, r, t)$ and $\boldsymbol{\mu}$ encodes the direction and magnitude of change from \textit{h} to \textit{t} via relation \textit{r}.
\par

The training uses the \gls{kg2e}\tilda\cite{ge-kg2e}, where each triple is evaluated to represent entities and relations in vector space accurately. Open world assumption (OWA) is used to prepare the dataset for the training algorithm. The KG dataset for the learning algorithm is defined as the positive triples $F^{+} = (h, r, t)$ and negative triples $F^{-} = (h', r,' t')$. Here, the negative triples are constructed by randomly choosing head and tail entities from the positive triples.\par

\subsection{The Proposed Intent Processing Pipeline}
\begin{figure*}[!htbp]
\begin{center}
    \includegraphics[width=0.8\textwidth]{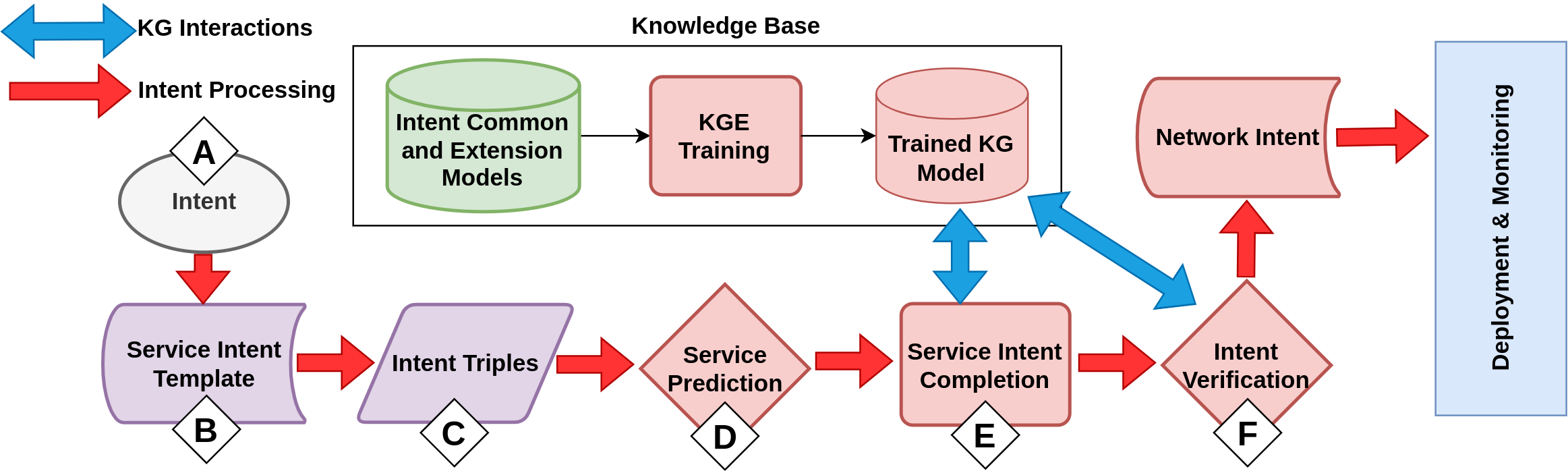}
\end{center}
\caption{Proposed intent processing pipeline with \gls{kge}.}
\label{image: ibn-fwork}
\end{figure*}

We next present the pipeline for utilizing \gls{kge} learning in the intent processing life cycle with active control of the deployed services (\figurename\tilda\ref{image: ibn-fwork}). 

The different stages in the translation and mapping of user intents to orchestrated service deployment are described as follows:
\setlist{nolistsep}
\begin{itemize}[noitemsep]
    \item Intent expression and recognition (\textbf{A});
    \item Creation of service intent template (\textbf{B});
    \item Identification of incomplete intent triples (\textbf{C});
    \item Service predictions from trained \gls{kge} model (\textbf{D});
    \item Completion of service intent template (\textbf{E});
    \item Verification of intent as network intent (\textbf{F}).
\end{itemize}

Step \textbf{A} involves identifying and extracting named entities from the user's intent, which delineates the search parameters and scope for subsequent phases of intent translation. Subsequently, Step \textbf{B} creates an intent template structured according to a hierarchical model, leveraging the intent ontology stipulated by \gls{tmf}\tilda\cite{tmforum-intent-mgmt-ont-tr292}. However, missing details concerning available resources, mapped service offerings, and \glspl{kpi} remain absent from the service intent template. Step \textbf{C} is devoted to identifying incomplete triples within the service intent, requiring supplementation with service-specific data. Step \textbf{D} employs link prediction techniques facilitated by a trained \gls{kge} model to infer resource, parameter, and KPI values. The link prediction task provides a set of \textbf{p} predictions with the highest scores. The contextual information from the \gls{ner} and the list of predictions are used to populate the missing information in the intent triples. In Step \textbf{E}, the predicted entities from the KG embedding are incorporated and expanded within the service intent, generating an unverified rendition of the network intent. Although comprehensive, this state of the intent necessitates verification, accomplished through a triple classification procedure using the pre-trained \gls{kge} model. Finally, Step \textbf{F} entails intent verification, whereby modified triples are evaluated based on the scoring function defined in Eq. (\ref{eq:3}). The intent verification is a triple classification task within the KGE modeling. It classifies the complete intent triples from the service intent as valid or invalid based on the pre-defined scoring function. At this point, the validated network intent encapsulates all requisite information for deploying the requested service, as ascertained by the intent processing pipeline.

\section{Performance Evaluation}
The intent processing pipeline is implemented using \textit{python} and the \textit{pykeen}\footnote{https://github.com/pykeen/pykeen} package. In addition, the process of \gls{ner} is implemented using \textit{spacy}\footnote{https://spacy.io/api/entityrecognizer} library with a custom word corpus consisting of key terms relevant to the context and expression style of the intent stakeholders. A public version of the intent processing pipeline is available via github\footnote{https://github.com/kashifme224/kg-embedding-pykeen}.

\subsection{\gls{kge} Model Training (KG2E)}
\begin{figure}[!ht]
    \includegraphics[width=0.45\textwidth]{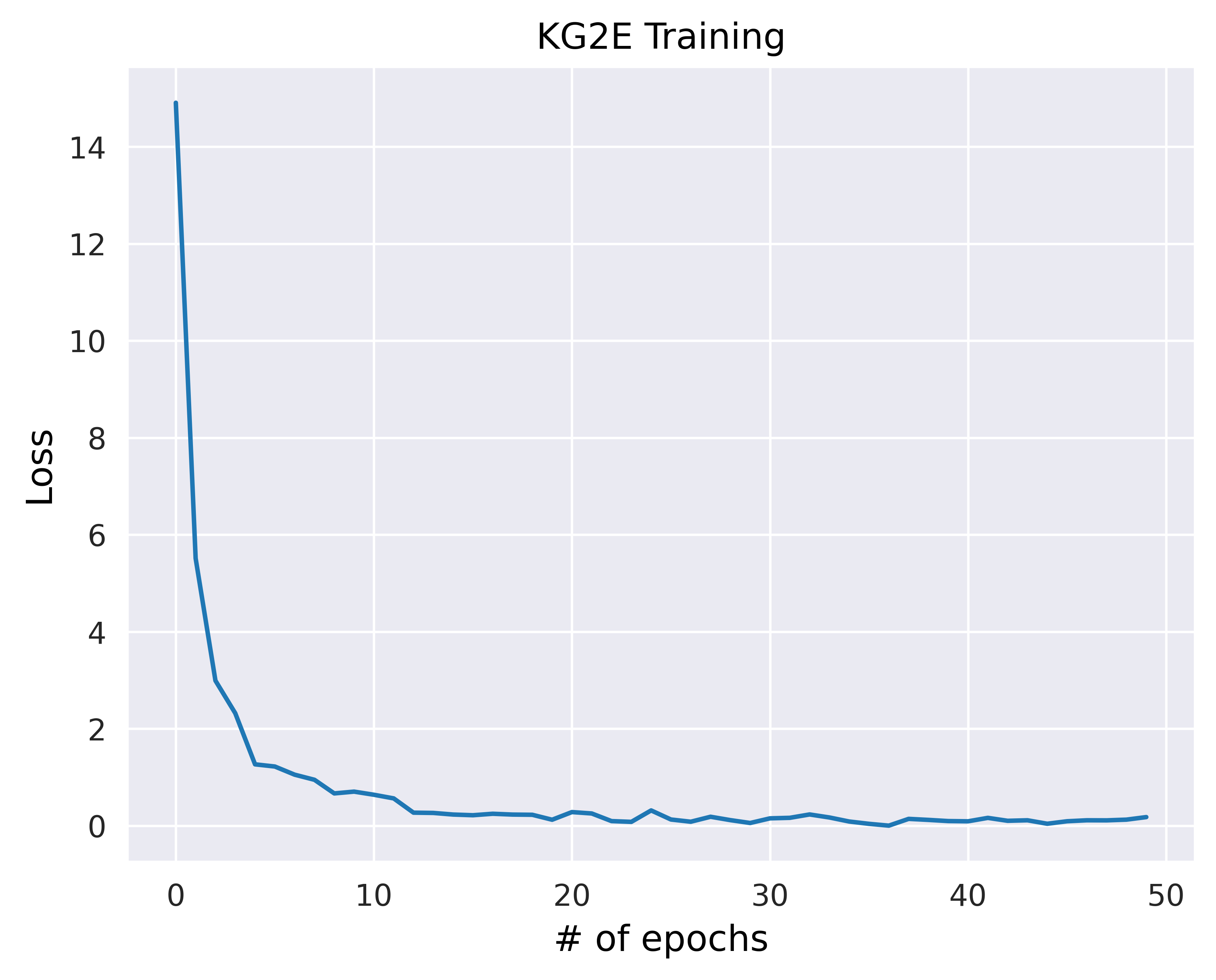}
    \caption{KG2E Model Training Convergence}
    \label{res: kge_loss}
\end{figure}

The training for the \gls{kge} model is performed with 50 epochs and an adaptive learning rate. In addition, this model uses a root mean square optimizer and calculates the distance measure based on KL-divergence. A total of 1575 triples (both negative and positive) are utilized for training with a (0.8, 0.1, 0.1) split for train, test, and validation sets.  The performance of the training loss minimization is depicted in \figurename ~\ref{res: kge_loss} with convergence after 12 optimization epochs.


\setlength{\belowcaptionskip}{0pt}
\begin{table}
\centering
\caption{Trained \gls{kge} model performance.}
\label{tab: kge_metrics}
\begin{tabular}{@{}lclc@{}}
\toprule
\multicolumn{2}{c}{\textbf{Link Prediction Metrics}} & \multicolumn{2}{c}{\textbf{Classification Metrics}} \\ \cmidrule(lr){1-2} \cmidrule(lr){3-4} 
\multicolumn{1}{l}{\textbf{\# of triples}}   & 1575  & \multicolumn{1}{l}{\textbf{accuracy}}    & 71.66    \\ 
\multicolumn{1}{l}{\textbf{Mean Rank}}      & 4.5     & \multicolumn{1}{l}{\textbf{f1\_score}} & 0.9830  \\ 
\multicolumn{1}{l}{\textbf{hits@1}}  & 0.58823 & \multicolumn{1}{l}{\textbf{TPR}}       & 0.75722 \\ 
\multicolumn{1}{l}{\textbf{hits@3}}  & 0.76470 & \multicolumn{1}{l}{\textbf{TNR}}       & 0.98301 \\ 
\multicolumn{1}{l}{\textbf{hits@5}}  & 0.82352 & \multicolumn{1}{l}{\textbf{FPR}}       & 0.24271 \\ 
\multicolumn{1}{l}{\textbf{hits@10}} & 0.91176 & \multicolumn{1}{l}{\textbf{FNR}}       & 0.01698 \\ \bottomrule
\end{tabular}%
\vspace{-0.6cm}
\end{table}
\setlength{\belowcaptionskip}{-10pt}

\begin{figure*}[!htbp]
\begin{center}
    \includegraphics[width=\textwidth]{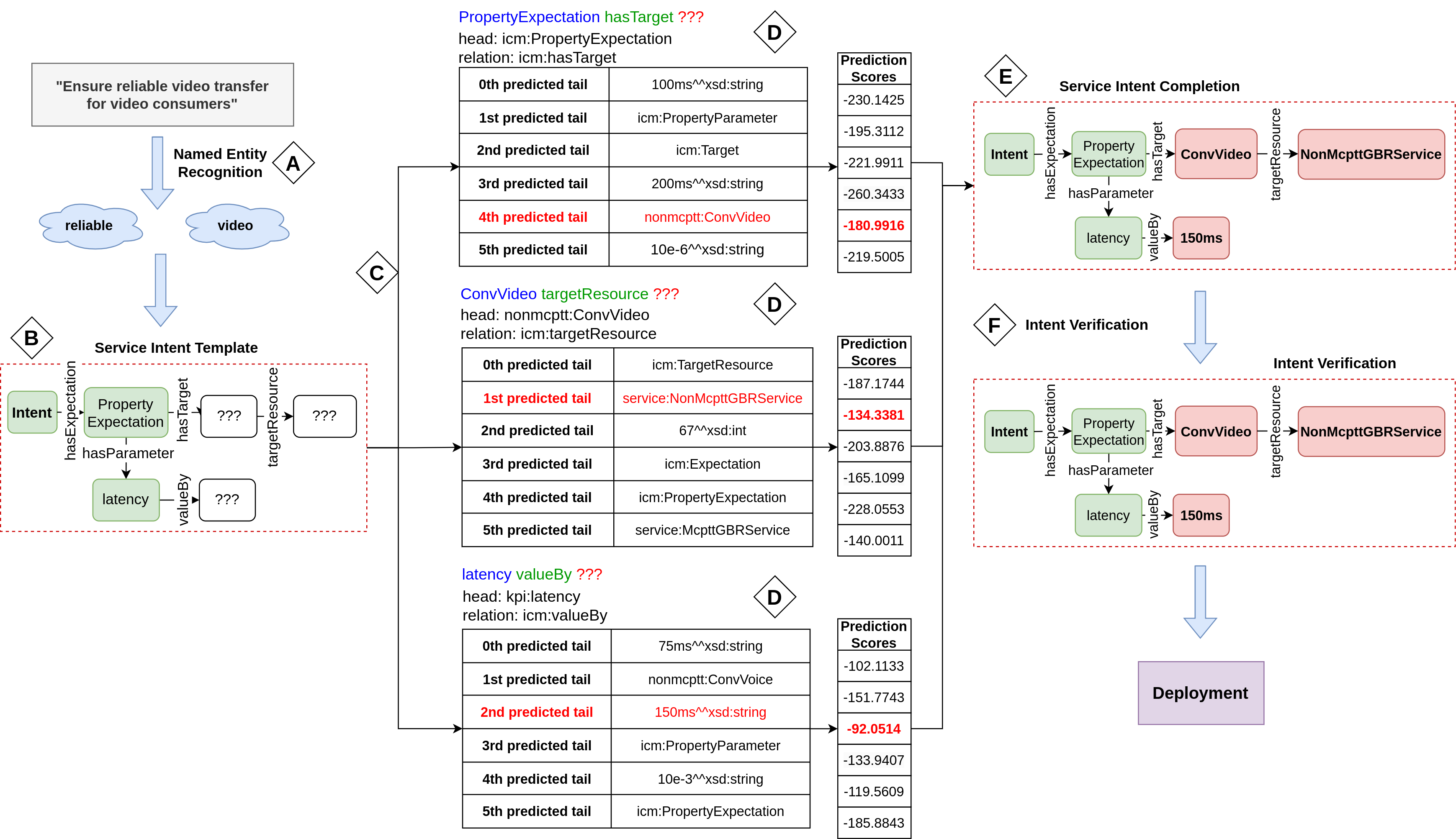}
\end{center}
\caption{Service intent completion and validation with link prediction and triple classification.}
\label{res: si-completion}
\end{figure*}

\subsection{Trained KGE Model Performance}
The performance of the trained \gls{kge} model for service predictions for intent-to-service mappings during the intent translation process is evaluated using typical \gls{kge} metrics\tilda\cite{kge-metrics} as shown in Table\tilda\ref{tab: kge_metrics}.
\subsubsection{Service Predictions}
We use \gls{kge} link predictions to perform service predictions and measure their performance with a rank-based evaluator. The rank-based evaluation protocol performs two link prediction tasks: 1) \textbf{right-side:} a pair of the head entity and relation is used to predict the tail entity \textit{(h,r,?)}, 2) \textbf{left-side:} a pair of the tail entity and relation is used to predict the head entity \textit{(?,r,t)}. The scores are calculated using the scoring function in Eq. (\ref{eq:3}) for both prediction tasks, and the scores are sorted in the order of decreasing score to determine the rank of the true choice (from the \gls{kg}). The rank is the index in the sorted list of scores. The evaluation metrics for rank-based analysis are as follows: 1) \textbf{Mean Rank}: the average rank of the correct entities and relations, 2) \textbf{Hits@p}: the proportion of valid entities or relations ranked in top \textbf{p} predictions. Each service prediction task provides a list of candidate \textbf{p} predictions to complete the incomplete triples. These predictions are analyzed according to their score and given context for an intent to select a valid completion combination for the incomplete triples. A good embedding model should be able to have a low mean rank and a high proportion of hits for a given value of \textbf{p}. 

\subsubsection{Intent Verification}
We use \gls{kge} model's triple classification protocol to verify the validity of the intents using a classification-based evaluator. We first define basic terms for the performance metrics in our current scenario as:
\setlist{nolistsep}
\begin{itemize}[noitemsep]
    \item True positive (TP): A correctly predicted intent as valid is also valid in the KG.
    \item True negative (TN): A correctly predicted intent as invalid is also invalid in the KG.
    \item False positive (FP): An incorrectly predicted intent as valid is invalid in the KG.
    \item False negative (FN): An incorrectly predicted intent as invalid is valid in the KG.
\end{itemize}
This type of evaluation performs the classical sensitivity and specificity analysis for prediction tasks using a standard set of metrics such as 1) Accuracy: ratio of the number of correct classifications to the total number, 2) f1-score: harmonic mean of precision calculated as $\sfrac{TP}{(TP+FP)}$ and recall calculated as $\sfrac{TP}{(TP+FN)}$, 3) \Gls{tpr}: the probability of correctly predicting a valid intent, 4) \Gls{tnr}: the probability of correctly predicting an invalid intent, 5) \Gls{fpr}: the probability of incorrectly predicting a valid intent, 6) \Gls{fnr}: the probability of incorrectly predicting an invalid intent. 




\subsection{An Example Intent Processing and Translation Flow}
An example flow of the intent processing pipeline is shown in \figurename\tilda\ref{res: si-completion}. The process starts with identifying service keywords (\textit{`reliable' and `video'}) from the natural language user intent expression (Step \textbf{A}). A blueprint service intent template is created and populated with relevant complete and incomplete triples from the \gls{ikg} (Step \textbf{B}). Afterward, the recognized service keywords are used for getting and selecting relevant service predictions from the trained \gls{kge} model for service intent completion (Steps \textbf{C, D}). The service intent completion (Step \textbf{E}) produces a service intent with all triples in a completed state. Intent verification classifies the service intent as valid or invalid based on the \gls{ikg} (Step \textbf{F}).

\subsubsection{Service Intent Completion}
Incomplete triples are supplemented with missing tail entries by leveraging the available head and relation entities. For instance, the triple \textit{(icm:PropertyExpectation, icm:hasTarget, \textbf{???})} identifies its pertinent prediction of \textit{nonmcptt:ConvVideo} having the highest prediction score (\textbf{-180.9916}) from the pool of predictions. Each incomplete intent triple determines the service, resource, and parameter necessary for completing the service intent template. The hierarchical predictions for each component are determined based on the scoring function delineated in Eq.~(\ref{eq:3}). Furthermore, the intent verification process (Step \textbf{F}) within the intent processing pipeline enables the filtering of suitable predictions from the prediction pool.\par

\subsubsection{Verified Network Intent}
The integration of the intent processing and translation pipeline is enhanced by incorporating an intent verification step, facilitated by the concept of triple classification\tilda\cite{know-graph-survey} utilizing a trained \gls{kge} model. To illustrate, consider the scenario depicted in \figurename\tilda\ref{res: si-completion}, where the incomplete triple \textit{(nonmcptt:ConvVideo, icm:targetResource, \textbf{???})} is supplemented by the top-ranked prediction \textit{(service:NonMcpttGBRService)} with the highest prediction score (\textbf{-134.3381}). However, it is worth noting that the prediction \textit{(service:McpttGBRService)} also warrants validation, as it possesses the second-highest prediction score (\textbf{-140.0011}) among the available recommendations, and it is also a valid completion for the incomplete intent triple. Conversely, the remaining predictions fail to adhere to the fundamental ontology of the knowledge graph, rendering them ineligible for validation. Consequently, an alternate prediction is sought to ensure the selection of a valid intent triple completion. Finally, the incomplete triple \textit{(kpi:latency, icm:valueBy, \textbf{???})} is completed with the top-ranked prediction (\texttt{150ms\string^\string^xsd:string}) with the highest score of (\textbf{-92.0514}). The resultant completed and verified service intent is referred to as the network intent, per the terminology established by the \gls{tmf}\tilda\cite{tmforum-intent-common-model-tr290}, encompassing all requisite service parameters for deployment within the underlying network infrastructure.

\section{Conclusion}
In this paper, we have explored the utilization of \gls{kge} and its potential application for intent processing and translation for \gls{ibn}. We have highlighted the significance of \glspl{kg} in capturing and representing complex network intents, service dependencies, and resource relationships. The incorporation of \gls{kge} models facilitates the extraction of meaningful insights from these knowledge graphs, enabling accurate prediction and completion of intent triples within the \gls{ibn} pipeline. 




%
\bibliographystyle{ieeetr}

\bibliography{refs}

\end{document}